# Rejuvenating the Contract of Academia With Society

Academia owes the public a fresh look at its education and research mission

———

By Abraham Loeb on February 6, 2018

It is easy for professors to hide deep inside the trenches of their expertise, bunkers that protect them from criticism by "non-experts" and allow them to promote their ego without supervision. True, academic freedom is a precious commodity that should be held sacred to enable discovery of the truth in the face of sociological forces and ideological dogma. But academia also owes an explanation of what it does and why it is relevant for the rest of humanity. There is a risk, after all, that some experts will dedicate their lifetime to exploring how many angels can stand on a pin or what are the properties of regions of the multiverse that we will never be able to probe.

First and foremost, researchers must communicate the results of their latest studies in a truthful and meaningful way. This has the dual role of educating the public of the latest advances as well as documenting that federal funds and education benefits are being spent in a way that resonates with the taxpayer's interests. The illusion that students and the educated public get from reading scholarly books is that the truth is revealed through an ordered, logical and pedagogical procedure. The reality is that research does not resemble a pristine landscape but rather a battlefield with lots of casualties (both from enemy and friendly fire), miscalculations, wrong assumptions or conclusions, and sociological intrigue. The path to finding the truth is often convoluted; competition is fierce and both experiments and theories could be wrong. Some argue that it is better to hide this dirty laundry from the public's eyes. For example, when the BICEP2 signal of the cosmic microwave background was first interpreted as evidence for gravitational waves from the beginning of the universe and then understood to be emission from interstellar dust, some of my colleagues worried that the public will now stop believing in the robustness of scientific predictions such as those involving climate change. But on the other hand, recognizing the reality of research would cultivate an atmosphere conducive for innovation in which the public accepts some level of risk of failure in view of the great benefits that come along with ground-breaking discoveries. The accelerating advances of science and technology leave no doubt that in the long run it is worth taking risks and accepting failure.

Second, the traditional boundaries among disciplines should be blurred since innovation often blossoms along these boundaries. Universities should consider a new organizational structure that moves away from the existing system of departments and enables a continuum of expertise across the arts, humanities and sciences. Students should be encouraged to take courses in multiple disciplines and organically weave them into new research patterns.

Third, universities should develop courses which are relevant for the skills required in the job market today. This means updating the courses every few years to accommodate new trends on topics ranging from artificial intelligence and big data to alternative energy sources or genome editing.

Overall, professors should mentor the future leaders of science, technology, arts and humanities and not just replicate themselves. We tend to like what we see when looking in the mirror, and so we often define our life mission to replicate scholars in our own image with expertise identical to ours. There is no doubt that this approach advances the popularity of our particular research program and promotes longevity of our ideas and style of thought. The louder the sound of consensus in the echo chamber of academia, the greater the boost to the ego of the people who constructed this chamber in the first place. But there is no guarantee that a loud voice speaks the truth better than a soft voice, and so diversity of opinion, which comes together with diversity of gender and ethnic origin, is always healthier for innovation and progress. Awarding prizes and societal memberships to those who maintain dominance of a united mainstream view should be substituted by rewarding innovation and encouraging independent thought. The promotion of diversity of ideas implies accepting the legitimacy of opposing opinions and criticisms. This does not mean that all opinions are equal but rather that alternative opinions should be discussed and filtered based on merit and clear reasoning as we iterate towards a better future.

I often find my conversations with laypersons more illuminating and inspiring than those with my professional colleagues. My wife often finds me speaking for hours with workers who fix the plumbing or landscape the yard at our home. For too long universities have been engaged in a monologue advocating their privileges. It is time for us to engage once again in a dialogue with society.

## ABOUT THE AUTHOR

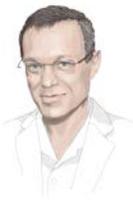

**Abraham Loeb**

Abraham Loeb is chair of the astronomy department at Harvard University, founding director of Harvard's Black Hole Initiative and director of the Institute for Theory and Computation at the Harvard-Smithsonian Center for Astrophysics. He serves as vice chair of the Board on Physics and Astronomy of the National Academies and chairs the advisory board for the Breakthrough Starshot project.

Credit: Nick Higgins